

\font\titolino=cmbx10
\font\tsnorm=cmr10
\font\tscors=cmti10
\font\tsnote=cmr7

\font\tscorsp=cmti9
\magnification=1200
\hsize=148truemm
\hoffset=10truemm
\parskip 3truemm plus 1truemm minus 1truemm
\parindent 8truemm

\def\2d{two-\-di\-men\-sio\-nal }
\def\3d{three-\-di\-men\-sio\-nal }
\def\4d{four-\-di\-men\-sio\-nal }

\def\PRD{{\tscors Phys. Rev. D }}
\def\NPB{{\tscors Nucl. Phys. B }}

\def\CMP{{\tscors Commun. Math. Phys. }}

\def\mpl{M_{pl}^2}
\def\d{\partial}

\def\sc{\scriptstyle}
\def\scc{\scriptscriptstyle}
\newcount\notenumber

\def\note{\advance\notenumber by 1 \footnote{$^{\the\notenumber}$}}
\def\ref#1{\medskip\everypar={\hangindent 2\parindent}#1}
\def\beginref{\begingroup
\bigskip
\leftline{\titolino References.}
\nobreak\noindent}
\def\endref{\par\endgroup}
\def\ra{\rightarrow}
\def\beginsection #1. #2.
{\bigskip
\leftline{\titolino #1. #2.}
\nobreak\noindent}

\nopagenumbers
\null
\vskip 5truemm
\rightline {INFNCA-TH9501}
\rightline {SISSA 16/95/A}
\rightline{February 1995}
\vskip 15truemm
\centerline{\titolino INSTABILITY OF THE $\bf R^3\times S^1$ VACUUM}
\bigskip
\centerline{\titolino IN LOW-ENERGY  EFFECTIVE STRING THEORY}
\vskip 15truemm
\centerline{\tsnorm Mariano Cadoni$^{(a),(c)}$ and Marco
Cavagli\`a$^{(b)}$}
\bigskip
\centerline{$^{(a)}$\tscorsp Dipartimento di Scienze Fisiche,}
\smallskip
\centerline{\tscorsp Universit\`a  di Cagliari, Italy}
\bigskip
\centerline{$^{(b)}$\tscorsp Sissa - International School
for Advanced Studies,}
\smallskip
\centerline{\tscorsp Via Beirut 2-4, I-34013 Trieste, Italy.}
\bigskip
\centerline{$^{(c)}$\tscorsp INFN, Sezione di Cagliari,}
\smallskip
\centerline{\tscorsp Via Ada Negri 18, I-09127 Cagliari, Italy.}
\bigskip
\vskip 15truemm
\centerline{\tsnorm ABSTRACT}
\begingroup\tsnorm\noindent
We present and discuss an euclidean solution of the low--energy
effective
string action that can be interpreted as a semiclassical decay
process
of
the ground state of the theory.
\vfill
\leftline{\tsnorm PACS: 11.25.-w, 11.25.Mj, 04.20.Jb\hfill}
\smallskip
\hrule
\noindent
\leftline{E-Mail: CADONI@CA.INFN.IT\hfill}
\leftline{E-Mail: CAVAGLIA@TSMI19.SISSA.IT\hfill}
\endgroup
\vfill
\eject
\footline{\hfill\folio\hfill}
\pageno=1
In Ref. [1] the authors found an instanton solution of a
four--dimensional, modulus field dependent, low--energy effective
string theory. That solution describes either a wormhole connecting
two
asymptotically flat regions or the nucleation of a baby universe
starting
from an original flat region. Our aim here is to show how this
instanton
can also describe a different physical process taking place in the
theory.
Indeed, using a different analytical continuation to the hyperbolic
space,
the solution of Ref. [1] can be interpreted as a semiclassical decay
process of the ground state (vacuum) of the theory.
The existence of a process of semiclassical decay is important since
it may lead to the instability of the vacuum of the theory.
Furthermore, a careful analysis of the geometric and topological
features of the instanton will enable us to identify the
wormhole solution of Ref. [1] as an Hawking--type wormhole [2]
connecting
two asymptotic regions of $R^3\times S^1$ topology.
The fact that an euclidean instanton could be interpreted either
as a wormhole or as a vacuum decay process is a very important and
appealing
new result.

In this paper we will follow an approach similar to the one
used by Witten in ref. [3] to prove the semiclassical
instability  of the Kaluza--Klein vacuum in five dimensions.
Even though the theory considered here has little to do with the
Kaluza--Klein theory in five dimensions, both instantons have common
geometrical and topological features and consequently most of the
mathematical techniques used in [3] can also be implemented in our
case.

Our starting point is the euclidean action ($(16\pi
G)^{-1}\equiv\mpl/16\pi=1$):
$$\eqalign{S_E=\int_\Omega d^4x\sqrt{|g|}e^{-2\phi}
\biggl[-R&+{\sc{8k}\over\sc{1-k}}
(\nabla\phi)^2+\varepsilon{\sc{3+k}\over\sc{1-k}}F^2\biggr]\cr\cr
&-2\int_{\d\Omega}
d^3x \sqrt{h}e^{-2\phi}({\bf K}-{\bf K_0})\,,}\eqno(1)$$
where $R$ is the curvature scalar, $\phi$ is the dilaton field,
$F_{\mu\nu}$ is the usual electromagnetic (EM) field tensor
and $k$ is a coupling constant, $-1\le k\le 1$. The boundary term is
required by
unitarity [4] and it is necessary whenever one deals
(as in our case) with asymptotically flat spaces; ${\bf K}$ is
the trace of the second fundamental form ${\bf K}_{ij}$ of the
boundary
 and
$\bf{K_0}$ is that of the asymptotic three--surface embedded in flat
space. $\varepsilon=\pm 1$ is a parameter whose meaning will be clear
in a moment.

Action (1) follows from the modulus--dependent low--energy effective
string theory considered in [5] once one eliminates the modulus from
the action by choosing an appropriate ansatz consistent with the
field
equations [6,1]. The action describes a Jordan-Brans-Dicke theory
coupled
to the electromagnetic field and reduces to well known theories
according
to the value of $k$: for instance, for $k=-1$ (1) coincides with the
usual low--energy string action when there is no modulus field [7,8];
for
$k=0$ the two--dimensional reduction of the action (1) gives the
Jackiw--Teitelboim theory [6,9]; finally, using some calculation
subtlety,
for $k=1$ we recover the usual Einstein-Maxwell theory.

The meaning of the parameter $\varepsilon$ needs some further
explanation.
As shown in [1], in order to write the contribution of the EM field
to
the
Lagrangian in a space with signature $(+,+,+,+)$ we have to choose
the
sign
of the term $F^2$ according to the electric or magnetic configuration
of
the field. Indeed, the EM field in the euclidean space is not
analytically
related to the EM field in the hyperbolic space by the simple
transformation
$t\ra i\tau$, but in general we have:
\medskip\noindent
\line{\hfill$E^2_{\rm hyp}=\varepsilon E^2_{\rm
Eucl}\,,$\hfill $H^2_{\rm hyp}=-\varepsilon H^2_{\rm
Eucl}\,.$\hfill (2)}
\medskip\noindent
So, keeping as usual $\varepsilon=1$, a real magnetic field in the
hyperbolic spacetime does not give, once continued in the euclidean
space, a real field. Since we wish to deal with real analytical
continuations of (real) hyperbolic fields in the euclidean space, we
allow for a different sign in front of the $F^2$ term in the action,
according to the configuration of the EM field. We will choose
$\varepsilon=-1$ for a purely magnetic configuration and
$\varepsilon=1$
for a purely electric one.\note{\tsnote Also duality invariance
arguments
support this prescription (see [1] for details). These arguments are
similar to those used in Ref. [10] for the case of the axion. The
key point is that $\sc F\to{}^{\scc *}F$ and the continuation to the
euclidean space do not commute.}

Now, let us consider a four--dimensional Riemannian manifold
described by a line element of the form:
$$ds^2=A^2(r)dt^2+B^2(r)d\chi^2+r^2d\Omega_2^2\,,\eqno(3)$$
where $\chi$ is the coordinate of the one-sphere, $0\le\chi<2\pi$,
and
$d\Omega_2^2=d\theta^2+\sin^2\theta d\varphi^2$ represents the line
element of the two-sphere $S^2$. Choosing for the EM field the
magnetic
monopole configuration on $S^2$ (and thus $\varepsilon=-1$)
$$F=Q_m\sin{\theta}d\theta\wedge d\varphi\,,\eqno(4)$$
the solution of the field equations derived from (1) is
$$\eqalignno{&ds^2=\left(1-{Q^2\over r^2}\right)^{-1}dr^2
+Q^2\left(1+{Q\over r}\right)^{k-1}
\left(1-{Q^2\over r^2}\right)d\chi^2+r^2d\Omega^2\,,&(5)\cr\cr
&e^{2(\phi-\phi_0)}= \left(1+{Q\over r}\right)^{(k-1)/2}
\,,&(6)\cr\cr}$$
where the magnetic charge $Q_m$ has been redefined through
$$Q_m={1\over 2}\sqrt{1-k} \, Q\,.\eqno(7)$$
The crucial point for the identification of (4-6) with a vacuum decay
process
is the analytical continuation of the line element to the hyperbolic
space.
Therefore, let us discuss the geometric and topological properties of
the
euclidean manifold described by (5). Since the latter has by
definition
signature $(+,+,+,+)$, $r$ can take values only in the range
$[Q,\infty[$.
For $r\to\infty$ the space is asymptotically flat with topology
$R^3\times S^1$. For $r=Q$ the metric tensor is singular. However, in
$r=Q$
the manifold is smooth, as it can be shown putting
$r=\sqrt{Q^2+\tau^2}$
($\tau\in]-\infty,\infty[$) and defining $\chi$ as a periodic
variable
with period $2\pi\cdot 2^{1-k}$ [1].
This conclusion seems to indicate that the coordinate system
$(r,\chi,\theta,\varphi)$ does not cover the whole manifold. In order
to
obtain the maximal extension of the euclidean metric (5) we have to
perform
an appropriate coordinate transformation:
$$\eqalign{&r=Q\cosh\left[\ln{\sqrt{x^2+t^2}\over
Q}\right]={(x^2+t^2)+Q^2\over 2\sqrt{x^2+t^2}}\,,\cr\cr
&\tan\theta ={x\over t}\,.}\eqno(8)$$
The inverse of (8) is:
$$\eqalign{&x=f(r)\sin\theta\cr\cr
&t=f(r)\cos\theta\cr\cr}\eqno(9)$$
where
$$f(r)=\sqrt{x^2+t^2}=Q\exp[\hbox{arccosh~}(r/Q)]\,.\eqno(10)$$
The coordinate transformation (9) is never singular. Using (8) the
euclidean
solution (4-6) becomes:
$$\eqalignno{&ds^2={1\over 4}\left(1+{Q^2\over f^2}\right)^2
\left[dt^2+dx^2+x^2d\varphi^2\right]+&\cr\cr
&\qquad\qquad +Q^2\left(1-{2Q^2\over f^2+Q^2}\right)^2
\left(1+{2Qf\over f^2+Q^2}\right)^{k-1}d\chi^2\,,&(11)\cr\cr
&e^{2(\phi-\phi_0)}=
\left(1+{2Qf\over f^2+Q^2}\right)^{(k-1)/2}\,,&(12)\cr\cr
&F={1\over 2}\sqrt{1-k}Q{x\over f^3}\left[x~dt\wedge
d\varphi~-~t~dx\wedge d\varphi\right]
\,.&(13)\cr\cr}$$
Eq. (11) represents the maximal extension of (5). As before, when
$x,t\to\infty$ the manifold is asymptotically flat with topology
$R^3\times S^1$. The critical surfaces are two: $x^2+t^2=Q^2$ and
$x^2+t^2=0$. Using the coordinate transformation it is easy to verify
that the first one corresponds to $r=Q$. The second critical surface
corresponds to $r=\infty$. The origin of the $(x,t)$ plane represents
thus a second asymptotically flat region. We have the situation
illustrated
in Fig. 1: two asymptotically flat regions smoothly joined through
the
circumference of radius $Q$. This strange structure is related to the
existence of
a conformal equivalence between the region inside $x^2+t^2=Q^2$ and
the
region outside. In fact, the euclidean line element (11) is invariant
under
the transformation:
$$y^\mu\to {Q^2\over y^2}O^\mu{}_\nu
y^\nu\,,\qquad\qquad\mu=1,2,3\eqno(14)$$
where $y^\mu$ are cartesian coordinates of the three--dimensional
space
$(t,x,\varphi)$, $y^1=t$, $y^2=x\cos\varphi$, $y^3=x\sin\varphi$, and
$O^\mu{}_\nu$ is a $3\times3$ rotation matrix. The region I is mapped
by
(14) in the region II, and vice versa. Hence, solution (11)
represents
a
Hawking--type wormhole [2] with minimum radius equal to $Q$
connecting
two asymptotically flat spaces with topology $R^3\times S^1$.
Note that (14) is an invariance of the entire solution (11-13) not
only of the metric (11). Indeed, also the expressions (12,13) for the
dilaton and the EM field do not change under the transformation
(14).

How can we recover the vacuum decay interpretation? In order to
answer
to
this question, we have to go  back to (5) and continue analytically
the
euclidean solution to a hyperbolic spacetime. In Ref. [1] the
analytical
continuation was performed first by defining $\tau=\sqrt{r^2-Q^2}$
thereafter by the complexification of $\tau$, $\tau\to i\tau$. The
resulting
hyperbolic manifold was interpreted as a baby universe of spatial
topology
$S^2\times S^1$ nucleated at $\tau=0$. However, the latter is not the
only
analytic continuation we can perform. For instance, we can complexify
the $\theta$ coordinate of the two--sphere $S^2$. In this case, since
$\theta=0$ is a coordinate singularity of the metric, it is
convenient
to
choose as symmetry plane the surface $\theta=\pi/2$ and to put
$$\theta\to {\pi\over 2}+i\xi\,.\eqno(15)$$
After the replacement (15) we obtain the hyperbolic solution:
$$\eqalignno{ds^2=\left(1-{Q^2\over r^2}\right)^{-1}dr^2
+Q^2\left(1+{Q\over r}\right)^{k-1}
&\left(1-{Q^2\over
r^2}\right)d\chi^2+\cr\cr
&-r^2d\xi^2+r^2\cosh^2\xi
d\varphi\,,&(16)\cr\cr
e^{2(\phi-\phi_0)}= \left(1+{Q\over r}
\right)^{(k-1)/2}&\,.&(17)\cr\cr}$$
The EM 2--form is now:
$$F=Q_m\cosh{\xi}d\xi\wedge d\varphi\,.\eqno(18)$$
The EM field is real, due to the choice $\varepsilon=-1$ in the
action
(1). For $r\ge Q$ this spacetime is nonsingular, the coordinate
singularity
at $r=Q$ being harmless as it was for the euclidean space (5). The
solution (16) for $r\ge Q$ represents the spacetime in which the
$R^3\times S^1$ vacuum decays. The topology of the initial $\xi=0$
surface is $R^2\times S^1$. Note that the analytic continuation to
the
hyperbolic space of Ref. [1], even though it was obtained from the
euclidean istanton (5), has instead spatial topology $S^2\times S^1$.

The topology of the analytic continuation to the hyperbolic space
depends
thus on the coordinate chosen to complexify. A better understanding
of
the
features of this space can be achieved starting from a hyperbolic
line
element that covers only the region $r\ge Q$. Using the coordinate
transformation
$$\eqalign{&x=f(r)\cosh\xi\,,\cr\cr
&t=f(r)\sinh\xi\,,\cr\cr}\eqno(19)$$
where $f(r)=\sqrt{x^2-t^2}$ is defined as function of $r$ as in Eq.
(10), we obtain:
$$\eqalignno{&ds^2={1\over 4}\left(1+{Q^2\over f^2}\right)^2
\left[-dt^2+dx^2+x^2d\varphi^2\right]+&\cr\cr
&\qquad\qquad +Q^2\left(1-{2Q^2\over f^2+Q^2}\right)^2
\left(1+{2Qf\over f^2+Q^2}\right)^{k-1}d\chi^2\,,&(20)\cr\cr
&e^{2(\phi-\phi_0)}=
\left(1+{2Qf\over f^2+Q^2}\right)^{(k-1)/2}\,,&(21)\cr\cr
&F={1\over 2}\sqrt{1-k}Q{x\over f^3}\left[x~dt\wedge
d\varphi~-~t~dx\wedge d\varphi\right]
\,.&(22)\cr\cr}$$
Since $-1\le t/x\le 1$, the new coordinates $(x,t)$ do not cover the
whole plane. They cover only the region outside to the light cone
$x=\pm t$,
corresponding to the physical region. As for the euclidean case, the
critical surfaces are two: $x^2-t^2=Q^2$, corresponding to $r=Q$, and
$x^2-t^2=0$ representing the infinity (see Fig. 2). Of course, the
manifold described by (20) is {\tscors geodesically complete} and its
topology is $R^3\times S^1$. Regions I and II in Fig. 2 are analogous
to the euclidean ones in Fig. 1 and their conformal equivalence can
be
proved
using a coordinate transformation similar to (14).

The region II is the starting point for the vacuum decay
interpretation
of the euclidean instanton. As one can easily verify, the
origin of the euclidean plane $(x,t)$ -- coinciding with an
asymptotically flat infinity -- is not the only surface we can use
to perform the analytic continuation in the hyperbolic space.
At $t=0$ we can join the euclidean manifold
described by (11) with a hyperbolic spacetime, namely the region
$x^2-t^2>Q^2$ of the spacetime (20)  (region II in Fig. 2). Indeed,
at
$t=0$
the metric, the dilaton field and the EM field assume a minimal
configuration, so the extrinsic curvature vanishes and the joining is
possible. The hyperbolic spacetime in which the vacuum decays is
the region II in Fig. 2. Let us explore in detail its properties. Due
to
the maximal analytic extension, the regions on  the left and on the
right
of the plane $(x,t)$ are identical, so we will focus our attention to
one of them. Choosing for simplicity $\chi=constant$, the line
element
(20) becomes conformally equivalent to a $R^3$ flat minkowskian
spacetime. Of course
the manifold is not geodesically complete, since there exist
geodesics
crossing the boundary $x^2-t^2=Q^2$. The meaning of the boundary can
be
understood following its time evolution. Starting at $t=0$, as
$t$ becomes larger and larger, the coordinate $x$ of the boundary
grows
 according to $x=\sqrt{Q^2+t^2}$.
Since the coordinate $x$ corresponds to a radius in the cylindrical
system of coordinates $(t,x,\varphi)$, the boundary can be
interpreted
as a
hole in space starting with radius $Q$ at $t=0$ and growing up for
$t>0$.
At $t=0$
the EM field is a  purely electric field in the $\varphi$
direction $E_\varphi=Q_m/x$; as the time $t$ flows and $E_\varphi$
changes in intensity, the latter generates a magnetic field in the
perpendicular $\chi$ direction.  Finally, when $x,t\to\infty$, the EM
field
vanishes, as expected because the spacetime is asymptotically flat.
The
euclidean line element (11) represents thus the decay process of the
flat spacetime of topology $R^3\times S^1$ in a spacetime with a
growing
hole.

In conclusion, the euclidean instanton we are dealing with represent
either  a wormhole or a vacuum decay process according to the
null--extrinsic
curvature surface used for the analytic continuation
to the hyperbolic spacetime.

The previous results can be straightforwardly extended to the purely
electric EM field configuration. Choosing $\varepsilon =1$ in the
action (1) and using an electric field along the $\chi$ direction
we have the solution:
$$F={1\over 2}\sqrt{1-k}Q^2 e^{2\phi_0}{1\over r^2}dr\wedge d\chi\,,
\eqno(23)$$
$$\eqalignno{&ds^2=e^{4\phi_0}\left(1-{Q\over r}\right)^{1-k}
\left[\left(1-{Q^2\over r^2}\right)^{-1}dr^2+\right.\cr\cr
&\qquad\left. +Q^2\left(1+{Q\over r}\right)^{k-1}
\left(1-{Q^2\over
r^2}\right)d\chi^2+r^2d\Omega^2\right]\,,&(24)\cr\cr
&e^{2(\phi-\phi_0)}= \left(1+{Q\over r}\right)^{(1-k)/2}
\,.&(25)\cr\cr}$$
Since the metric (24) differs from the previous one for the purely
magnetic
case only through a conformal factor, all conclusions remain
unchanged.

At this stage we can ask ourselves if the semiclassical vacuum decay
process
is consistent with energy conservation. Since the $R^3\times S^1$
vacuum has
zero energy, the space (16) in which it decays must also have zero
energy.
Using the ADM formula generalized to dilaton--gravity theories, the
total
energy of (16--18) can be calculated as usual by means of
a surface integral depending
on the asymptotic behaviour of the gravitational and dilaton fields.
The line element (20) is not static with respect to $t$, so the
integral
must be evaluated at the initial $t=0$ surface, corresponding in (16)
to
$\xi=0$. The result of the integration is zero.
Indeed, the terms of the gravitational and
dilaton fields which contribute to the total energy of the solution
are
those of order $1/r$. However, in our case these terms give a null
contribution to the energy, owing to the $R^2\times S^1$ topology of
the
$\xi=0$ surface. The space described by (16) has therefore zero
energy.
This feature makes the $R^3\times S^1$ vacuum not stable for the
theory defined by (1), since there exists a solution with zero energy
and
the same asymptotic behaviour as the $R^3\times S^1$ vacuum.
An important consequence of this result
is that the positive energy theorem [11] does
not hold for the theory (1) if one considers vacua with topology
$R^3\times S^1$.
The positive energy theorem states that every non-flat,
asymptotically
minkowskian solution of the Einstein equations has zero energy.
However,
its validity for spaces with arbitrary topology and for theories as
(1) is
difficult to prove. In the case under consideration the failure of
the
positive energy theorem seems related to the presence of the EM
field:
in
the $R^3\times S^1$ vacuum there exist excitations of the EM field
for
which
the positive energy theorem does not hold.

The interpretation of the euclidean solution (5) as an instability of
the
vacuum has been established using the analytical continuation (15).
Considering a second analytical continuation to a hyperbolic
spacetime,
we have also seen that the instanton can be interpreted as a
Hawking--type
wormhole. The latter has an intrinsically three--dimensional nature
because its
topology is $R^3\times S^1$ and the radius of $S^1$ is equal to $Q$
in
the two asymptotic regions $f=\infty$, $f=0$ and shrinks to zero
for $r=Q$. Hence, the most natural interpretation of this solution
can
be
found in the context of a $3+1$ Kaluza--Klein theory in  which,
compactifying a dimension, we end up with a three--dimensional
gravity
theory plus a scalar field that parameterizes the compactified
dimension.

Starting from the action (1) with $\varepsilon=-1$, setting to zero
the components of the EM field along the $\chi$ direction
and splitting the four--dimensional line element as
$$ds^{(4)}=ds^{(3)} +Q^2 e^{-2\psi}d\chi^2\,,\eqno(26)$$
after some manipulations we obtain the three-dimensional action:
$$S_E=\int_\Omega d^3x\sqrt{|g^{(3)}|}e^{-2\sigma}
\biggl[-R^{(3)}+{\sc{k-1}\over\sc{2}}
[4(\nabla\sigma)^2-(\nabla \eta)^2]-{\sc{3+k}\over
\sc{1-k}}F^2\biggl]\,,
\eqno(27)$$
where $\sigma=\phi +\psi/2$, $\eta=\psi+2\phi(k+1)/(k-1)$, and we
have
dropped the boundary terms.

A solution of the ensuing equations of motion is:
$$\eqalignno{&ds^2={1\over 4}\left(1+{Q^2\over f^2}\right)^2
\left[dt^2+dx^2+x^2d\varphi^2\right]\,, &(28)\cr\cr
&e^{2(\sigma-\sigma_0)}=
{f^2+Q^2\over f^2-Q^2}\,,&(29)\cr\cr
&e^{2(\eta-\eta_0)}=
\biggl({f+Q\over f-Q}\biggr)^2
\,,&(30)\cr\cr}$$
where $f=\sqrt{x^2+t^2}$  and we have chosen the EM tensor $F$ as
in (13). The solution of the three--dimensional theory is thus a
Hawking--type wormhole connecting two asymptotic regions of topology
$R^3$.

Now, let us calculate the decay rate of the vacuum.
Evaluating the action (1) on the euclidean solution (4--6) we have
$$S_E=4\pi^2e^{-2\phi_0} Q^2(k+1)\,.\eqno(31)$$
This result has been obtained by integrating $r$ and $\theta$
 in the range
$Q\le r<\infty$, $0\le\theta\le \pi/2$, the appropriate one
for the vacuum decay process.
The rate of decay of the $R^3\times S^1$ vacuum is:
$$\Gamma_{vd}=\exp\biggl[- 4\pi^2e^{-2\phi_0}
Q^2(k+1)\biggr]\,.\eqno(32)$$
The vacuum is long--lived for values of $Q$ much greater than the
Planck
length $l_p$ and becomes unstable when $Q$ is of the same order of
magnitude
of $l_p$. Finally, it is interesting to compare the vacuum decay rate
$\Gamma_{vd}$ with the probability for the nucleation of a baby
universe
$\Gamma_{\sc{bu}}$ (see Ref. [1]):
$$\Gamma_{\sc{bu}}=(\Gamma_{\sc{vd}})^2\,.\eqno(33)$$
Hence, the probability of nucleation of a baby universe is smaller
than the probability of the vacuum decay.
\vfill\eject
\null
\beginref

\ref [1] M. Cadoni and M. Cavagli\`a, \PRD {\bf 50}, 6435 (1994).

\ref [2] S.W. Hawking \PRD {\bf 37}, 904 (1988).

\ref [3] E. Witten, \NPB {\bf 195}, 481 (1982).

\ref [4] S.W. Hawking, {\tscors in:} ``General Relativity, an
Einstein Centenary Survey'', eds. S.W. Hawking and W. Israel
(Cambridge University Press, Cambridge, 1979).

\ref [5] M. Cadoni and S. Mignemi, \PRD {\bf 48}, 5536 (1993).

\ref [6] M. Cadoni and S. Mignemi, \NPB {\bf 427}, 669 (1994).

\ref [7] D. Garfinkle, G.T. Horowitz and A. Strominger, \PRD
{\bf 43}, 3140 (1991).

\ref [8] G.W. Gibbons and K. Maeda, \NPB {\bf 298}, 741 (1988).

\ref [9] M. Cadoni and S. Mignemi \PRD (in press), hep-th/9410041.

\ref [10] S.B. Giddings and A. Strominger, \NPB {\bf 306}, 890
(1988).

\ref [11] P. Schoen and S.T. Yau, \CMP {\bf 65}, 45 (1979).

\endref
\bye